\documentclass[12pt,preprint]{aastex}
\usepackage{emulateapj5,apjfonts}

\newcommand{\EXO}{\mbox{EXO 0748-676}}

\newcommand{\Fe}{${\rm Fe} \ $}

\shorttitle{Neutron Star Redshift Measurements} 
\shortauthors{Bildsten, Chang and Paerels}

\begin{document}

\title{Atomic Spectral Features During Thermonuclear Flashes on 
Neutron Stars}

\author{Lars Bildsten\altaffilmark{1}, Philip Chang\altaffilmark{2},
and Frits Paerels\altaffilmark{3}}
\altaffiltext{1}{Kavli Institute for Theoretical Physics,  Kohn Hall, University
of California, Santa Barbara, CA 93106, USA; email:
bildsten@kitp.ucsb.edu}
\altaffiltext{2}
{Department of Physics, Broida Hall, University of California,
Santa Barbara, CA 93106; pchang@physics.ucsb.edu}
\altaffiltext{3}
{Columbia Astrophysics Laboratory and Department of Astronomy,
Columbia University, 538 W. 120th St., New York, NY 10027;
frits@astro.columbia.edu}

\begin{abstract}

The gravitational redshift measured by Cottam, Paerels and Mendez for
the neutron star (NS) in the low-mass X-ray binary EXO~0748-676
depends on the identification of an absorption line during a type I
burst as the H$\alpha$ line from hydrogenic Fe. We show that Fe is
present above the photosphere as long as $\dot M>4\times
10^{-13}M_\odot {\rm yr^{-1}}$ during the burst. In this limit, the
total Fe column is $N_{\rm Fe}\approx 3\times 10^{19}{\rm cm^{-2}}$
for incident material of solar abundances and only depends on the
nuclear physics of the proton spallation. The Fe destruction creates
many heavy elements with $Z<26$ which may imprint photo-ionization
edges on the NS spectra during a radius expansion event or in a burst
cooling tail. Detecting these features in concert with those from Fe
would confirm a redshift measurement. We also begin to address the
radiative transfer problem, and find that a concentrated Fe layer with
$kT=1.2-1.4\ {\rm keV}$ and column $N_{\rm Fe}= 7-20 \times 10^{20}\
{\rm cm}^{-2}$ (depending on the line depth) above the hotter
continuum photosphere is required to create the H$\alpha$ line of the
observed strength. This estimate must be refined by considerations of
non-LTE effects as well as resonant line transport. Until these are
carried out, we cannot say whether the Fe column from accretion and
spallation is in conflict with the observations.  We also show that
hydrogenic Fe might remain in the photosphere due to radiative
levitation from the high burst flux.

\end{abstract}

\keywords{diffusion -- nuclear reactions -- 
stars: abundances, surface -- stars: neutron -- X-rays: binaries, bursts}

\section{Introduction}

  The dominant thermal flux during type I bursts and superbursts offer
an ideal opportunity for studying neutron star (NS) properties through
spectral measurements of their photospheres (e.g. London, Taam \&
Howard 1984, 1986; Foster, Ross \& Fabian 1987). Though never
confirmed, the reports by Waki et al. (1984), Turner \& Breedon
(1984), Nakamura, Inoue \& Tanaka (1988) and Magnier et al.  (1989) of
4 keV absorption lines during type I bursts motivated theoretical work
on the formation of the Fe Ly$\alpha$ line and edge (e.g. Foster et
al. 1987; Day, Fabian \& Ross 1992). Cottam, Paerels \& Mendez (2002;
hereafter CPM) observed the accreting NS \EXO\ using the Reflection
Grating Spectrometer (RGS) aboard the XMM-Newton satellite and claimed
detection of hydrogen-like and helium-like Fe absorption features from
the $n=2\rightarrow 3$ H$\alpha$ transitions during type I bursts.
This {\it Letter} is an initial study on the implications of the
hydrogen-like feature and predictions for heavy element abundances
near the photosphere. The helium-like feature will be left for future
work.

XMM/Newton observed \EXO\ for 335 ks during which 28 type I X-ray
bursts were seen. CPM combined the burst spectra and split them into a
high luminosity ($L$) part (from the initial rise and fall) and a low
$L$ part, which covers the fading to the baseline $L$. In the high $L$
part, the detected absorption line was identified as the H$\alpha$
transition of hydrogenic Fe. In the low $L$ spectrum, a line
identified as the He-like H$\alpha$ transition was seen. For these
identifications, CPM measured the NS's gravitational redshift at
$1+z=1.35$, giving $R=4.4GM/c^2$ from the line centroid (neglecting
the rotational corrections of Fujimoto 1985; \"Ozel \& Psaltis 2003).
These spectra have photospheric temperatures $kT_{\rm eff,
\infty}\approx 1.6 \ {\rm keV}$ and $kT_{\rm eff, \infty}< 1.4 \ {\rm
keV}$, using the measured color temperatures ($kT_C=1.8 \ {\rm keV}$
and $kT_C<1.5 \ {\rm keV}$; CPM) and correcting for the redshift and
spectral hardening from Comptonization (London et al. 1986; Madej
1997).

In \S~2, we calculate the Fe abundance when accretion is active,
showing that proton spallation of Fe (Bildsten, Salpeter \& Wasserman
1992; hereafter BSW) naturally yields $N_{\rm Fe} = 3.4 \times
10^{19}\ {\rm cm}^{-2}$ for accretion of solar metallicity
material. The spallation also produces large columns of elements with
$Z<26$, which could be observable via their photoionization edges
during the cooling phase of the burst or a radius expansion burst.  In
\S~3, we show that Stark broadening (Paerels 1997) dominates the
H$\alpha$ line equivalent width and that the measured equivalent width
(properly redshifted to the NS surface) of $W_{\rm H\alpha}=10$ eV
implies an Fe column of $N_{\rm Fe} \approx 7-20 \times 10^{20}\ {\rm
cm}^{-2}$ depending on line depth. However, these abundance
calculations are highly uncertain due to non-LTE (NLTE) effects and
resonant line transport which we believe play an important role in the
line formation.  We summarize in \S 4 and note that the radiative flux
during the burst can levitate hydrogenic Fe.

\section{Heavy Element Abundances from Accretion}

The continuum photosphere for the 1.276 keV photons of the Fe
H$\alpha$ transition is at an electron column density of $N_{\rm
e}\approx \sigma_{\rm Th}^{-1}$, where $\sigma_{\rm Th}$ is the
Thomson cross section. If \Fe is uniformly abundant at the solar value
($f_i \approx 3\times 10^{-5}$ by number; Grevesse \& Sauval 1999)
above the continuum photosphere at 1 keV then $N_{\rm Fe}\approx
f_i/\sigma_{\rm Th}\approx 5\times 10^{19}\ {\rm cm^{-2}}$. Joss
(1977) showed that convection during the burst cannot bring burning
fluid elements to the photosphere as the entropy there exceeds that in
the burning layers by large factors. Hence, we look to Fe in the
accreted material as the likely source of the photospheric
elements. BSW studied the deposition and separation of accreting
elements, and we use their results here.

Imagine first that accretion halts during the burst.  In a pure H
atmosphere, the upward pointing atmospheric electric field is
$m_pg/2$, so that the force on Fe is $F=-43m_pg$. The
counter-balancing drag force on the drifting Fe nucleus is
$F_{dr}=Z^2r_{ep}(m_p/m_e)^{1/2}\rho v/m_p$, where $r_{ep}$ is the
electron-proton coupling from BSW.  The downward drift speed is
$v\approx 1 \ {\rm cm \ s^{-1}}T_7^{3/2}/\rho$ so that it takes
$t_{\rm s}\approx 1 \ {\rm s}(y/1\ {\rm g \ cm^{-2}})/T_7^{3/2}$ for
Fe to fall to a column depth $y$.  Hence, if accretion halts, Fe would
sink beneath the photosphere in much less than the burst duration.

Given the sub-Eddington peak luminosity, accretion likely remains
active during the bursts.  Kuulkers et al. (2002) have shown this to
be the case for GX 17+2.  The persistent flux from \EXO \ when type I
bursts occur implies $\dot M> 2\times 10^{-10} M_\odot {\rm yr^{-1}}$
(Gottwald et al. 1986; CPM).  In order to obtain a lower limit on the
accretion rate per unit area we assume that the accretion covers the
whole NS, so that $\dot m> 10^3 {\rm \ g \ cm^{-2}\ s^{-1}}$.  The
measured redshift implies $R=4.4GM/c^2$ (CPM), so that the NS's
surface is inside the last stable orbit regardless of its rotation
rate (Kluzniak \& Wagoner 1985; Kluzniak, Michelson \& Wagoner
1990). The resulting accretion gap allows for the accreting beam to
strike the atmosphere with a velocity comparable to free-fall
(Kluzniak \& Wilson 1991), realizing the Coulomb stopping scenario of
Zel'dovich and Shakura (1969). The integration of a free trajectory
from the last stable orbit to the surface yields an impact angle
$\sin\theta\approx 0.1$ and kinetic energy of $\approx 200-300$
MeV/nucleon.

Since some angular momentum and energy can be lost in the accretion
gap, we presume that $\sin\theta>0.1$ and fix the incident energy at
$200$ MeV/nucleon, resulting in stopping a particle of charge $Ze$ and
mass $Am_p$ after traversing a vertical column $y_{\rm s}\approx
5.4\sin\theta (m_p/\sigma_{\rm Th})A/Z^2$. BSW showed the
repercussions of the $A/Z^2$ scaling: (i) the incident electrons stop
first, (ii) hydrogen and helium penetrate furthest to $y_{\rm s,p}$,
and (iii) all heavier elements stop prior to the H/He beam. For
example, Fe stops at $y_{\rm s,Fe}=0.08y_{\rm s,p}$ and is spalled by
the energetic protons (of current $J_p\approx \dot m/m_p$) until it
has sunk to safety (i.e.  $y>y_{\rm s,p}$). This is the mechanism
shown by BSW to destroy accreted heavy elements. Grazing incidence
reduces the distance that must be fallen to reach safety and increases
the surviving fraction of the incident CNO elements relative to that
in BSW.

 The infalling Fe could have bound electrons as it spirals into the NS
surface. However, the combination of varying Doppler shifts and the
short flight time compared to the residence time in the atmosphere
(derived below) means that the inflow is an unlikely environment for
narrow line formation. We thus focus our discussion on the thermalized
Fe in the atmosphere.  At 200 MeV per nucleon, $\approx 30$\% of the
incident Fe suffer a charge changing nuclear reaction before
traversing $\approx 1\ {\rm g \ cm^{-2}}$ and thermalizing. Accretion
is thus depositing thermal Fe into the NS atmosphere at $y_{\rm s,
Fe}=1.0\sin\theta\ {\rm g \ cm^{-2}}$ at a rate $\dot N_{\rm
Fe}\approx 2f_i J_p/3$. For the spallation cross section (i.e. that
cross section for $^{56}{\rm Fe}$ to be chipped to lower Z)
$\sigma_D\approx 600 \ {\rm mb}$ (Silberberg et al. 1998) the Fe
lifetime to spallation from the proton beam is $t_{\rm d}=(\sigma_D
J_p)^{-1}\approx 2.8\ {\rm ms}(10^3\ {\rm g \ cm^{-2} \ s^{-1}}/\dot
m) $. The Fe must sink to a depth of $y_{\rm s,p}=14\sin\theta\ {\rm g
\ cm^{-2}}$ for safety, so that when $\dot m> 2\ {\rm g \ cm^{-2} \
s^{-1}}(0.1/\sin\theta)T_7^{3/2}$ (or $\dot M>4\times 10^{-13}M_\odot
\ {\rm yr^{-1}} T_7^{3/2}$), destruction dominates settling
(i.e. $t_{\rm s}>t_{\rm d}$).

 Hence, as long as $\dot M$ during the burst exceeds 1\% of that
occurring prior to the burst, the destruction by the proton beam
completely determines the Fe residence time in the upper atmosphere.
This seems a condition easily satisfied for many bursters and greatly
simplifies the calculation of $N_{\rm Fe}$ as well as the Stark
broadening discussed in \S~3, since, in this limit, the Fe is present
only near the stopping point. Because of the complicated interplay of
the currents and forces in the upper atmosphere (see BSW), the local
abundance near the stopping point ranges from $\approx 1-40$ times the
incident abundance. The proton spallation reduces the Fe abundance as
it sinks, imprinting an abundance profile $\propto \exp(-t_{\rm
s}/t_{\rm d})$, which is an exponential decay with depth. The Fe rich
region above the Fe poor layer puts a small inverted gradient in the
mean molecular weight ($\mu$) that would trigger a rapid overturn in
an isothermal layer if $|d\ln\mu/d\ln P|>2/5$. Gradients this large
will not be reached as long as the accreting material has an abundance
less than ten times solar. Doubly diffusive instabilites might still
occur, but those timescales are likely no shorter than $ t_{\rm s}$.

Hence, for finding $N_{\rm Fe}$, we only need to balance the Fe
destruction rate (neglecting the $\alpha$ beam is less than a 20\%
effect, as those cross sections are rarely larger than twice those
induced by protons; Ferrando et al. 1988) $\dot N_{\rm Fe}=N_{\rm
Fe}\sigma_{D}J_p$, with the Fe deposition rate $\dot N_{\rm
Fe}=2f_iJ_p/3$, giving
\begin{equation}
N_{\rm Fe}={2f_i\over 3\sigma_{D}}\approx 3.4\times 10^{19} \ {\rm
cm^{-2}}\left(f_i\over 3\times 10^{-5}\right)\left(600 \ {\rm mb}\over
\sigma_{D}\right). 
\end{equation}
Due to neutron knockout reactions, about 10\% of those Fe nuclei are
$^{55}{\rm Fe}$ and 5\% are $^{54}{\rm Fe}$. The Fe column is
comparable to the uniform solar abundance calculation because
$\sigma_D\approx \sigma_{\rm Th}$.

Since many spallation reactions occur as the nuclei sink, we also
found the abundances of all lower $Z$ and $A$ nuclei. We calculated
this nuclear cascade by following the evolution of each isotope under
the constraint that production equals destruction. The cross-sections
from Silberberg et al. (1998) were used.  We found that the sum of all
nuclei with $20<Z<26$ outnumber Fe by a factor of 2.5-3.0 (see Table
1). In the absence of the BSW spallation scenario, these nuclei would
not be present.  The single most abundant heavy nucleus is $^{56}{\rm
Fe}$.

 The immediate question that comes to mind is whether these spallation
products can be detected.  They may be detected via their H$\alpha$
lines, but the calculation of their predicted EW is left to future
work.  In Table 1, we find the $kT_{1/2}$ at which $1/2$ of these
nuclei would be in the hydrogenic state (for $y=0.5\ {\rm g\ cm^{-2}}$
in Table 1), allowing for a possible photoionization edge detection
once $kT <kT_{1/2}$ (presuming LTE).  To fit all $Z$, we used the
explicit cross-section from Rybicki \& Lightman (1979) and set the
bound-free Gaunt factor to $g=0.8$, giving $\sigma_{\rm
bf}(E)=6.32\times 10^{-18}\ {\rm cm^2}(E_{\rm e}/E)^3 Z^{-2}$ where
$E_{\rm e}$ is the edge energy (shown in Table 1, value in parentheses
is the redshifted edge energy) from Bethe \& Salpeter (1957; hereafter
BS). This $\sigma_{\rm bf}$ agreed to within (3-5)\% to Verner et
al.'s (1996) analytic fits for Ar, Ca and Fe. For now, we estimate the
impact of such an edge on the spectrum by considering the integral
over the optical depth above the photoionization edge, $\tau_{\rm
e}=N\sigma_{\rm bf}$, assuming an effectively cold atmosphere (i.e. no
reemission above the edge) $EW_{\rm e}=\int (1-\exp(-N\sigma_{\rm
bf}(E)))dE$. For $\tau_{\rm e}\ll 1$, the $EW_{\rm e}$ integral is
simply expanded to obtain $EW_{\rm e}=175\ {\rm eV}(N/10^{19}\ {\rm
cm^{-2}})$ integrated from $E=E_{\rm e}$ to $1.3E_{\rm e}$ which is
independent of $Z$.

Table 1 shows $EW_{\rm e}$ assuming the element is present at the
level implied by the spallation scenario and finding the amount of it
in the Hydrogenic ground state from Saha equilibrium for $kT=1.2 \
{\rm keV}$.  As the burst cools or during a radius expansion burst,
these edges could become prominent, and the entry of $\tau_{\rm e,max}
$ in Table 1 presumes all of the spallation created elements have at
least one bound electron. It is intriguing to note that these edge
energies and $EW_{\rm e}$ are close to those previously reported by
Waki et. al. (1984), Turner \& Breedon (1984), Nakamura et al. (1988),
Magnier et al.  (1989) and Kuulkers et al. (2002).

\section{Estimated Iron Abundances from Observed Equivalent Widths} 

We now estimate the Fe column needed to give the H$\alpha$ line
observed by CPM.  Motivated by our accretion scenario, we initially
presume that this line forms in a thin isothermal layer of Fe in LTE
at temperature $T$ above the hotter continuum photosphere.  The
observed redshift implies a surface gravity of $g\approx 3\times
10^{14}\ {\rm cm\ s}^{-2}\,(1.4M_\odot/M)$.  Though rapid rotation
($\approx 100\,{\rm Hz}$) might explain the observed line width
($\approx 40\,{\rm eV}$), the line's equivalent width depends most
strongly on 
the intrinsic broadening mechanisms. The scale for thermal Doppler
broadening is $\Delta E_D/E_0 = \sqrt{kT/Am_pc^2}\approx 1.5 \times
10^{-4} (kT/1\ {\rm keV})^{1/2}$.  The natural line width, $\Gamma
\approx 5\times 10^{13}\ {\rm s}^{-1}$ (BS), is given by the
spontaneous decay rate for the $3 \rightarrow 1,2$ transitions, which in
units of the thermal Doppler shift is $a = \hbar\Gamma/2\Delta E_D
\approx 0.3$.

We now show that Stark broadening is larger by an order of magnitude
than either of these effects for the H$\alpha$ transition.  Stark
broadening results from the shift of line energies due to the electric
field from neighboring ions, $e/r_0^2$, where $r_0 = (4\pi
n_e/3)^{-1/3}$ is the mean ion spacing (Mihalas 1978). The energy
perturbation from the linear Stark effect is $\Delta E_{\rm S} =e^2
\Delta x/r_0^2$, where $\Delta x$ is the size of the orbital.  Using
the linear Stark effect is justified since its energy shift at the
densities and temperatures considered exceeds that of the degeneracy
breaking Lamb shift for the relativistic H-like ion (BS). For the
$n=3$ state of Fe at a Thomson depth unity, we find ${\Delta E_{\rm
S}}/E_0 = 1.5 \times 10^{-3}$, an order of magnitude larger than
thermal Doppler.

Since the majority of the intrinsic line width is attributed to Stark
broadening, $N_{\rm Fe}$  must be found from the measured EW
given by the curve of growth (Mihalas 1978),
\begin{equation}
W = 2 A_0 \Delta E \int_0^{\infty} \frac {\eta(\alpha)} {1 +
\eta(\alpha)} d\alpha,
\end{equation}
where $W$ is the EW in energy units, $A_0$ is the normalized line
depth at line center, $\Delta E$ is the intrinsic energy width (in our
case the Stark shift), $\eta$ is the ratio of the line opacity to the
continuum opacity and $\alpha = E/\Delta E$. We use the electric field
distribution for a uniform ion sea (e.g. the Holtsmark distribution,
$W_H$; Mihalas 1978), giving $\eta(\alpha) = \eta_0 W_H(\alpha)$ where
$\eta_0 = {N_{\rm Fe, n=2} \sigma_{\rm line}}/{N_e \sigma_{\rm Th}}$
is the ratio of line optical depth to continuum optical depth at line
center. Depending on density either thermal Doppler or Stark
broadening dominates, so we convolve the Holtsmark distribution with
the Voigt profile to cover the complete range of densities.  We also
ignore Debye screening effects on the electric field distribution
since the deviation from the Holtsmark distribution is small for the
typical $T$ and $\rho$.

 Because rotational broadening acts to lower the apparent line depth
and is likely to be important, the observed value of $A_0\approx 0.3 $
is a lower limit. Indeed, the true line depth could be significantly
larger than 0.3. However, such a large value of $A_0$ is difficult to
understand for realistic temperature gradients and presuming LTE. For
a gray atmosphere (a reasonable approximation for the temperature
profile of a type I burst atmosphere) $A_0$ attains a maximal value of
0.2 when the Fe resides at very low column depths.  The resolution of
this dilemma may be the role of NLTE effects and resonant line
transport in the H$\alpha$ line. Resonant line transport could
generate such an absorption feature with a large line depth without
invoking large temperature contrasts as in LTE (Jefferies 1968). NLTE
effects clearly play a role in determining abundances from the EW
measurements.  For instance, an immediate measure of the role of NLTE
effects is that the radiative ionization rate for the Fe ground state
exceeds the collisional ionization rate (London et al. 1986),
complicating the ionization balance for Fe.

 To resolve the role of NLTE effects, the radiative transfer equation
must be solved self-consistently for both the resonant transport of
the H$\alpha$ and Ly$\alpha$ lines with detailed balanced of the Fe
ionization and excitation in the strong radiation field of a Type I
burst. For now, we simply present results for $A_0 \approx 0.3$, which
is set by observations and $A_0 = 0.2$, which is characteristic of a
gray atmosphere.

We calculate the $N_{\rm Fe}$ required to reach $W_{\rm H\alpha} = 10\
{\rm eV}$ presuming all Fe resides at the column depth, $y=P/g$.  We
assume LTE and that the dominant states for Fe are either fully
ionized, or in the hydrogenic ground state or helium-like ground state
to approximate the partition function.  For radial infall, with the Fe
layer between $kT = 1.2-1.4\ {\rm keV}$ and $A_0 = 0.3\ (0.2)$,
$N_{\rm Fe} = 7\ (20) \times 10^{20}\ {\rm cm}^{-2}$.  This
temperature range is consistent with the absence of helium-like Fe in
the high $L$ phase of the burst (CPM).  Since the occupation of the
$n=1$ state exceeds that in $n=2$ by $\sim 10^{2-3}$, we also
calculate $W_{\rm Ly\alpha}$ for the implied $N_{\rm Fe}$ values.  For
this line, thermal Doppler and natural line broadening dominate Stark
broadening (Madej 1989).  When $kT = 1.2 - 1.4\ {\rm keV}$ and the Fe
resides at the pressure implied by radial accretion, $W_{\rm Ly\alpha}
< 30\ {\rm eV}$.  Foster et al. (1987) also calculated $W_{\rm
Ly\alpha}$ for a uniformly distributed Fe abundance at the solar value
and found that it would not exceed $30\ {\rm eV}$. It would require
an $A_0\approx 1$ to give an Fe column consistent with that calculated
from spallation.  However, such large temperature contrasts (and hence
such large values of $A_0$) are not possible in these atmospheres and
we need to understand the NLTE line transport problem before we
can address this apparent inconsistency. 

 Day, Fabian \& Ross (1992) pointed out that a uniform presence of Fe
at the solar value would create an observable photo-ionization edge.
The Fe column require to produce the observed H$\alpha$ line would
yield enormous edges. However, NLTE effects would weaken the strength
of these photoionization features significantly (London et al. 1986).
The reported broad spectral features in the 6-9 keV range (van
Paradijs et al. 1990, Strohmayer \& Brown 2002) during type I bursts
and superbursts need to be re-analysed as possible Fe photo-ionization
edges.

\section{Summary and Conclusions}

  Active accretion in the context of a beam hitting the atmosphere
naturally gives $N_{\rm Fe}\approx 3\times 10^{19}\ {\rm cm}^{-2}$
which only depends on the physics of proton spallation of Fe. In this
scenario, heavy elements with $Z<26$ are produced in quantities
comparable to Fe and may be observed via their photoionization edges
or atomic lines (see Table 1) during the cooling phase of type I
bursts or superbursts. Our initial LTE calculations have found that an
Fe column of $\approx 7-20 \times 10^{20}\ {\rm cm}^{-2}$ is needed to
produce the measured $W_{\rm H\alpha}$ of CPM. However, until a more
complete calculation of the resonant line transport for the H$\alpha$
and Ly$\alpha$ line has been undertaken in this strong radiation
environment, we cannot say whether this factor of twenty discrepency
is certain.

 Throughout this paper, we have presumed that Fe is present because of
active accretion during the burst. However, the radiative force on
hydrogenic Fe in the ground state is $F_{\rm rad} = \int {\sigma_{E}
F_{E}}/{c}\ dE,$ where $F_{E}$ is the flux per photon energy at the
Ly$\alpha$ transition ($E=6.9\ {\rm keV}$). Since $F_{E}$ is nearly
constant across the line, $F_{\rm rad}= \pi h e^2 f_{1\rightarrow 2}
F_E/m_e c^2$ in the optically thin limit (Michaud 1970), where
$f_{1\rightarrow 2}=0.42$ is the oscillator strength from BS. If $F_E$
is given by a black-body, then at $kT_{\rm eff} = 1\ {\rm keV}$, the
radiative acceleration on Fe is $a_{\rm rad} \approx 10^{15} {\rm cm\
s}^{-2}$, a factor of three larger than $g$. The radiative force will
be even larger for realistic type I burst spectra, which are harder
than a blackbody at the observed $T_{\rm eff}$ (London et al. 1986;
Pavlov, et al. 1991).  The Fe diffusion time is less than the burst
duration, so there is adequate time for the accelerated Fe to reach a
new diffusive equilibrium. The calculation of the resulting Fe column,
which could well saturate the line and lower the radiative force, is
beyond the scope of this {\it Letter}, as it requires a simultaneous
solution of the radiative transfer and the Fe diffusion equations,
like that done for hot white dwarfs (e.g. Chayer, Fontaine \& Wesemael
1995).

\acknowledgments

 We thank A. Piro for calculating the infall energy and incident
angles and N. Barghouty for providing the spallation cross-section
code of Silberberg et al. (1998). We thank the referee, Marten van
Kerkwijk, for critical comments and Ira Wasserman for many insightful
conversation and for pointing out the critical role NLTE effects can
play in line formation. This work was supported by the National
Science Foundation under grants PHY99-07949 and AST02-05956 and NASA
through grant NAG 5-8658.  L. B. is a Cottrell Scholar of the Research
Corporation.

\begin{deluxetable}{c c c c c c}
  \tablecolumns{7} \tablewidth{0pt} \tablecaption{Heavy Element Abundances
  and Atomic Features}

  \tablehead{
    \colhead{} & \colhead{$E_{\rm e}$} &
    \colhead{$\tau_{\rm e,max}$} & \colhead{$N_i/N_{\rm Fe}$} &
    \colhead{$EW_{\rm e}$} & \colhead{$kT_{1/2}$}\\  
    \colhead{} & \colhead{[keV(redsh)]} &\colhead{}
    & \colhead{} & \colhead{[eV]} & \colhead{[keV]}}
  \startdata 
  Fe & 9.279(6.87) & 0.32 & 1.00 & 277 & 1.12\\ 
  Mn & 8.573(6.35) & 0.18 & 0.36 & 66 & 1.05\\ 
  Cr & 7.897(5.85) & 0.26 & 0.48 & 56 & 0.99\\ 
  V  & 7.248(5.37) & 0.26 & 0.43 & 32 & 0.93\\ 
  Ti & 6.627(4.91) & 0.31 & 0.48 & 22 & 0.86\\ 
  Sc & 6.035(4.47) & 0.41 & 0.58 & 17 & 0.81\\ 
  Ca & 5.471(4.05) & 0.47 & 0.59 & 11 & 0.75\\ 
  K  & 4.934(3.65) & 0.48 & 0.55 & 7 & 0.69 \\ 
  Ar & 4.427(3.28) & 0.53 & 0.54 & 4 & 0.65 \\ 
  \enddata
\end{deluxetable}
\end{document}